\documentclass[twocolumn,superscriptaddress,preprintnumbers,amsmath,amssymb]{revtex4}
\usepackage{graphicx}% Include figure files
\usepackage{dcolumn}% Align table columns on decimal point
\usepackage{stmaryrd}
\usepackage{array}
\usepackage{caption}
\usepackage{amsthm}
\usepackage{float}
\usepackage{braket}
\usepackage{bm}
\newtheorem{theo}{Theorem}
\newtheorem{csq}{Consequence}
\newtheorem*{prop}{Proposition}

\newcommand{\be}{\begin{equation}}
\newcommand{\ee}{\end{equation}}
\newcommand{\bea}{\begin{eqnarray}}
\newcommand{\eea}{\end{eqnarray}}

\newcommand{\sgn}{\mathrm{sgn}}
\newcommand{\card}{\mathrm{card}}
\newcommand{\PolyGamma}{\mathrm{PolyGamma}}
\begin{document}

%\title{ Exact evaluation of the correlation energy for an arbitrary
%        number of particles: the filling one quantum Hall state }

%\title{ Filling one integer quantum Hall state: exact results from
%        $N=2$ to $N \rightarrow \infty$ }

%\title{ Exact results for finite systems of electrons at 
%        filling one of the integer quantum Hall state: arbitrary $N$ }

%\title{ Exact results for finite systems of electrons in the 
%        integer quantum Hall phase with filling factor one: arbitrary $N$ }

%\title{ Exact results for finite quantum Hall systems of electrons at  
%        filling factor one: arbitrary $N$ }

\title{ On the distribution of charges in a conducting needle }

\author{Orion Ciftja}
%\email{ogciftja@pvamu.edu}

\affiliation{Department of Physics, Prairie View A\&M University,
             Prairie View, TX 77446, USA}

\author{Adrien Guénard}
\affiliation{adrien.guenard@free.fr}

\author{Nefton Pali}
\affiliation{Orsay Institut of Mathematics, University of Paris-Saclay, 307 Michel Magat, 91400 Orsay, France}

% \affiliation{Kavli Institute for Theoretical Physics,
%               University of California, Santa Barbara,
%               CA 93106, USA}

%\date{\today}
\date{September 18, 2025}

\begin{abstract}

We study the distribution of point charges in a straight conductive needle and the 
electric field created by them. Starting from the bead model with $n$ point charges on the needle, we show the existence and uniqueness of an equilibrium state. 
We also study the differential system pertaining to the system and 
show that the system of point charges does not converge towards an equilibrium state. 
In order to move from a discrete to a continuous model we increase the number of the 
point charges and explain the paradoxical convergence towards a uniform distribution 
although the charges tend to accumulate towards the ends of the needle 
(both at equilibrium and for the differential system of equations describing the 
Newtonian motion).
This convergence has to be understood as that of the distribution functions of a sequence of probabilities (whether it is at equilibrium or when the charges move). 
We also provide visual illustrations that help understanding the studied phenomena.

\end{abstract}

\maketitle

\section*{Introduction}

In 1878, Maxwell \cite{maxwell} had tackled the problem of the charge in 
a {``}long and narrow{''} cylinder, and had already raised the question of passing 
to the limit of needle. 
%
% Between 1878 and 1996,
Since then, different authors had studied the problem of 
charge distribution in a cylinder \cite{kapista,smythe}. 
As of recent times,
the problem of distribution of point charges in a needle was posed explicitly in 1996 
by D. J. Griffiths and Y. Li \cite{griffiths}. 
In their article, they provided several alternatives on how to approach the question, 
either by going to the limit starting from the charge on a cylinder or on a sphere, 
or by starting directly from the needle, within the framework of the bead model. 
%
% After Griffiths and Li's article, 
%
After this work, other physicists became interested in the problem. 
However, as M. Andrews pointed out in his 1997 article~\cite{andrews}, 
Griffiths and Li~\cite{griffiths} had not reached a definitive conclusion and 
various questions were still left open. 
They wondered in particular whether the density obtained by crossing the limit depended on the way in which this crossing was carried out. 
In 1997, Good \cite{good} gave a geometric interpretation of the problem based 
on the starting model of a sphere supporting the idea of a uniform distribution. 
J. D. Jackson carried out two exhaustive studies 
in 2000 and 2002 \cite{jackson2000,jackson2002} using limits from cylinders. 
In 2004, N. Amir and H. Matzner published an article \cite{amir} in which they also came to the conclusion of a uniform distribution, obtained from a study of the load on a cylinder. 
The surprising element of all these treatments is the fact that different models 
(for instance, an ellipsoid or a cylinder model) may lead to different results, thus,
there is even some ambiguity on whether the problem is well-posed.
A slightly different approach is that followed in 2017 by J. Batle et al~\cite{batle}
in which they did not consider the limit of any higher-dimensional body reducing to a needle.
Instead, they started with charges placed in a needle, but assumed that 
the charges interact with a non-Coulomb interaction potential
which reduces to the Coulomb one as a limit. 
This approach led to a uniform charge distribution being reached.
Desipte the reverence of this problem, in our opinion, 
the answer to the question of what is the 
equilibrium distribution of a given total charge contained 
on a straight needle of finite length does not have a fully unambigous answer?
In this work we revisit the problem from a much firmer grounded mathematical 
formalism as compared to earlier treatments that are more intutitive and physics-oriented.
We believe that this work gives a definitive answer for the mathematical 
solution of the problem of charge distribution in a needle.
The convergence towards a uniform charge distribution  
can be understood and can be properly interpreted only by seeing it 
as that of distribution functions of a sequence of probabilities.

\subsection*{Model and notations}

Hereafter, we approach the problem directly on a needle, using the bead model described in the article by Griffiths and Li. { }Here is what motivated this method: when looking at the problem from a microscopic point of view, the field is created by individual charges, and the continuous density is a convenient approximation of the discrete problem. As Griffiths and Li pointed out, starting from such an approximation and then taking another approximation, for example from a cylinder to the needle, could be more problematic than a direct approximation from the discrete case. 

For the study, we reduce this physics problem to a mathematical problem formulated using standard mathematical tools.

\textbullet\ The needle is brought back to the segment $[0,1]$, so that the set of possible positions became a well-known simplex of $\mathbb{R}^n$. 

\textbullet\ For the study of the field around the needle, we use $[0,1] \times \{ 0 \} \times \{ 0 \}$ in $\mathbb{R}^3$ to represent the needle.

\textbullet\ On the needle are $n$ identical charges, of value $q$, whose respective positions at time $t$ are $x_1(t)<x_2(t)< \cdots < x_n(t)$. Let us note, for all $i \in \llbracket 1,n \rrbracket$, $X_i(t)=\left( x_i(t), 0, 0 \right)$, and for all $u \in [0,1]$, $U=(u,0,0)$.

\textbullet\ We assume that the order between the charges is fixed. This allows us to number the charges. This is important to assure the uniqueness of the equilibrium point. The charges numbered $1$ and $n$ remain respectively at positions $0$ and $1$ on the needle. Indeed, they are pushed by the others charges
and we assume that they cannot escape from the needle.

\textbullet\ The field created by these charges at a point $X \in \mathbb{R}^3$ is $\frac{q}{4 \pi \varepsilon_0} \sum\limits_{i=1}^n \frac{1}{\left\| X-X_i \right\|_2^2} \left( \frac{X - X_i}{\left\| X-X_i \right\|_2} \right)$. The multiplicative factor $\frac{q}{4 \pi \varepsilon_0}$ not being involved in determining the equilibrium, we will omit it throughout the rest. Also, when modifying the number of charges on the needle, we will assume that the total needle load is $1$.

\textbullet\ By normalizing the total charge to $1$, we transform the charge distribution into a probability measure that gives us access to the usual convergence theorems.

\textbullet\ If at time $t$ the charges are in position $x_1(t)=0<x_2(t)< \cdots < x_n(t)=1$, the $i$-th charge undergoes from the other charges (up to the multiplicative coefficients) a force given by

\[  f_i(t)= \sum_{j=1}^{i-1} \frac{1}{\left(x_i(t)-x_j(t)\right)^2}-\sum_{j=i+1}^n \frac{1}{\left(x_i(t)-x_j(t)\right)^2}. \]

\textbullet\ The $n$ charges are at equilibrium position if for each $i\in \llbracket 2,n-1 \rrbracket$, $\sum\limits_{j=1}^{i-1} \frac{1}{\left( x_i-x_j \right)^2}=\sum\limits_{j=i+1}^n \frac{1}{\left( x_i-x_j \right)^2}$. 

\textbullet\ We study what happens when new charges are added. When we study the add of one charge to $n$ charges already in the needle, we note $x_1, x_2, \ldots , x_n$ the equilibrium positions of the initial $n$ charges, and $y_1, y_2, \ldots , y_{n+1}$ the equilibrium positions of the $n+1$ charges after one charge has been added.

\subsection*{Summary of results}

Due to the impossibility of an exact computation of the positions of the charges on the needle, we alternate extensive numerical simulations and theoretical considerations motivated by the results of the experimental study.

\textbullet\ We first show by an experimental study of a differential system that the charges on a needle move indefinitely around an average position.

\textbullet\ We then study the system of equations for an equilibrium position. The exact solution requires to find the roots of several variables polynomials of high degree. There is no way to solve this system by closed formulas. The solutions can only be approximated numerically. In order to prove the existence of the equilibrium, we use a purely mathematical method : we define a continuous function whose fixed points are exactly the equilibrium positions for the charges. The use of a fixed point theorem proves the existence of an equilibrium.

\textbullet\ We then use an argument of convexity in order to prove the uniqueness of the equilibrium. This provides another proof of the existence. The combination of the tools provided by these two proofs makes it possible to numerically calculate the equilibria.

\textbullet\ We then define another differential system which provides another way to estimate the equilibrium position, and give another proof of the existence of the equilibrium.

\textbullet\ We then study the evolution of the equilibrium position when adding a new charge. Extensive numerical simulations make appear a linear move at first order. We give by mathematical computations a more precise estimation of that first order approximation, and a return to experimental study confirms that estimation. This approximation has some consequences on the intervals between consecutive charges at equilibrium.

\textbullet\ We then give a precise mathematical meaning to the {``}convergence of the charge distribution towards an uniform charge density{''}. This solve the {``}paradox{''} pointed out by Griffiths and Li : convergence towards an uniform distribution although charges accumulate towards the ends of the
needle. For this, we use dyadic decompositions: after computing the equilibrium distribution for $2^n+1$ charges, $n \in \llbracket 2, 11 \rrbracket$, we can give both a visual explanation and a precise mathematical meaning to the convergence towards an uniform density. This can be done by a renormalization (total charge equal to 1) leading to a sequence of discrete measures of probabilities. \textbf{The convergence has to be understood as the convergence of the distribution functions of that sequence of probabilities.} 

\textbullet\ The convergence towards the uniform density having been explained when charges are at equilibrium, we come back to the differential system and check numerically that altough the charges move indefinitely and do not stay at equilibrium point, when the number of charges increases, the sequence of distribution functions still converge towards the distribution function of the uniform density. This is visually illustrated in the animated short film accompanying the article.

\textbullet\ We then study some consequences of the convergence of the distributions functions. Once again, using an appropriate theory of the integral, we can use a standard convergence theorem to prove that the sequence of fields induced by discrete distributions of charges converge towards the field induced by the continuous limit distribution defined above. By using a computation based on Cauchy{'}s principal value idea, we can extend the field function to the whole space (including the needle). 

\textbullet\ The theorem about the convergence of the distributions functions can be interpreted as a {``}principle of equivalence{''} : in order to study the field created by charges, one can modelize that by any repartition (discrete or continuous), as far as the distribution functions are close enough. Using that principle, we give some exact formulas for uniform discrete distributions of charges. 

\textbullet\ We finish by two graphical illustrations of the convergence of the fields : one as a function of the distance to the needle, and the other as a function of the number of charges.

\subsection*{Methods used}

\textbullet\ The study of equilibrium has been transformed into a fixed point problem thanks to the {``}erasing{''} of the singularities constituted by the charges and the use of a function using functions of the form $x \mapsto \mathrm{e}^{\frac{-1}{x^2}}$.

\textbullet\ We study two differential systems for which a special care of the domain is important. The numerical study is also delicate and requires high precision computations.

\textbullet\ The study of the convergence towards the limiting density was carried out by the use of dyadic numbers for the number of charges. This use of dyadic numbers made it possible to reveal the distribution functions of discrete densities. Returning the needle to $[0,1]$ allowed a more practical visual interpretation of the properties of the distribution functions obtained experimentally.

\textbullet\ The choice of an adapted integral theory (the Riemann-Stieltjes theory) made it possible to give a precise meaning to the convergence towards the limiting density (which made it possible to explain a {``}paradox{''} raised by Griffiths and Li on the behavior at the ends of the needle, and to answer a question on integrability). 

\textbullet\ Finally, the concrete calculations were carried out on \textit{Mathematica}.

\subsection*{Motivations}

The problem has been raised to us by the following note : 

Consider a conductive needle, say $[0,1] \times \{ 0 \} \times \{ 0 \}$ in $\mathbb{R}^3$, on which there are $n$ charges placed at positions $X_i = (x_i,0,0)$ ($i \in \llbracket 1,n \rrbracket$, $x_i \in [0,1]$). The field created by these charges at a point $X \in \mathbb{R}^3$ is $\frac{q}{4 \pi \varepsilon_0} \sum\limits_{i=1}^n \frac{1}{\left\| X-X_i \right\|_2^2} \left( \frac{X-X_i}{\left\| X-X_i \right\|_2} \right)$. If the number of charges increases, we are tempted to consider a density $t \mapsto \lambda (t)$ of charge on the needle, and to replace the sum by an integral, $\frac{1}{4 \pi \varepsilon_0} \int_0^1 \frac{1}{\left\| X-T \right\|_2^2} \left( \frac{X-T}{\left\| X-T \right\|_2} \right) \lambda (t) \mathrm{d}t$. If $X=(x,0,0)$ is a point on the needle, the field at $X$ becomes $\frac{1}{4 \pi \varepsilon_0} \int_0^1 \frac{1}{(x-t)^2} \sgn (x-t) \lambda (t) \mathrm{d}t$. The function $t \mapsto \frac{1}{(x-t)^2} \sgn (x-t) \lambda (t)$ is not defined at $x$, and the two integrals $\int_0^x \frac{1}{(x-t)^2} \lambda (t) \mathrm{d} t$ and $\int_x^1 \frac{1}{(x-t)^2} \lambda (t) \mathrm{d} t$ diverge unless $\lim\limits_x \lambda =0$ and therefore $\lambda (x)=0$. This having to be true for all $x \in [0,1]$, the only suitable density is zero.

Towards the end of this paper, we will explain what is wrong in this {``}proof{''}, and how is has to be modified in order to get the result.

\section*{1. Behavior of $n$ charges placed on a needle}

Before studying balance, let us see experimentally how charges behave; in particular if they tend to be distributed according to an equilibrium position.

Consider $n$ charges placed on the needle $[0,1]$, which we number from $1$ to $n$. As Griffiths and Li pointed out, charge n°$1$ is placed at $0$, and charge n°$n$ is placed at $1$. If at time $t$ the charges are in position $x_1(t)=0<x_2(t)< \cdots <x_n(t)=1$, the $i$-th charge undergoes from the other charges (up to the multiplicative coefficients) a force given by 
\[ f_i(t)= \sum_{j=1}^{i-1} \frac{1}{\left(x_i(t)-x_j(t)\right)^2}-\sum _{j=i+1}^n \frac{1}{\left(x_i(t)-x_j(t)\right)^2}. \]

To understand how loads behave over time, we considered the differential system

\begin{equation}
\begin{cases}
x_1 (t) =0&\\
x_n (t) = 1&\\
x_i'' (t) = f_i (t)&\text{if $1 < i < n$}\\
x_i (0) = \frac{i-1}{n-1}&\text{if $1 \leqslant i \leqslant n$}\\
x_i' (0) = 0&\text{if $1 \leqslant i \leqslant n$}
\end{cases}.
\label{systemone}
\end{equation}

For reasons that we will give later, we have numerically computed solutions of this differential system for $n = 2^p+1$, $p \in \llbracket 2, 7 \rrbracket$. In each case, with the exception of the charges numbered $1$, $2^{p-1}+1$ and $2^p+1$ which remain fixed, the other charges oscillate indefinitely around an average position which, as we will see later, is the equilibrium position. This {``}eternal{''} oscillations look like those obtained with the pendulum equation: with the elementary equation, one finds a pendulum which oscillate indefinitely. When the equation is modified with a damping term, the pendulum stabilizes progressively at the equilibrium point. A question is : how has the differential system of the charges on a needle to be modified?\\
Furthermore, as $p$ increases, the amplitude of the oscillations of each charge decreases. Here are the trajectories obtained for the first values of $p$ (There is no benefit to increase the number of charges: when this number is too high, one cannot see anymore the movement of the individual charges). We will see later how the charges behave when initial conditions are irregular.

$n=2^3+1=9$

\begin{figure}[H]
\begin{center}
\includegraphics[scale=0.5]{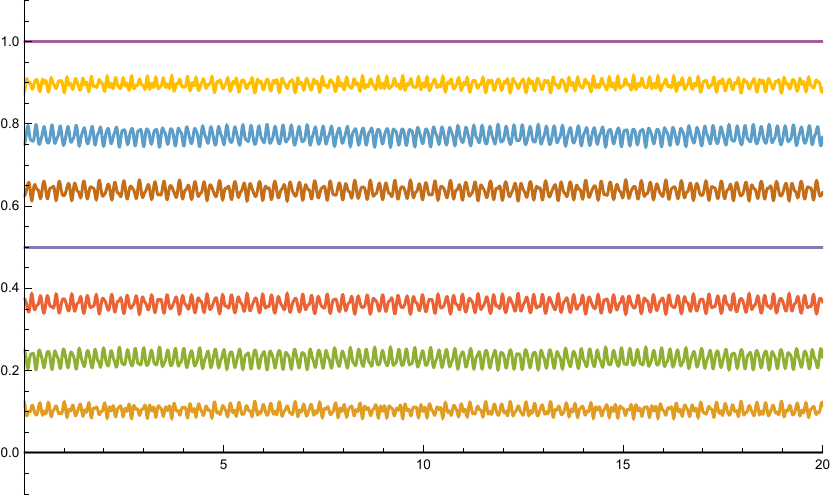}
\end{center}
\end{figure}

$n=2^4+1=17$

\begin{figure}[H]
\begin{center}
\includegraphics[scale=0.5]{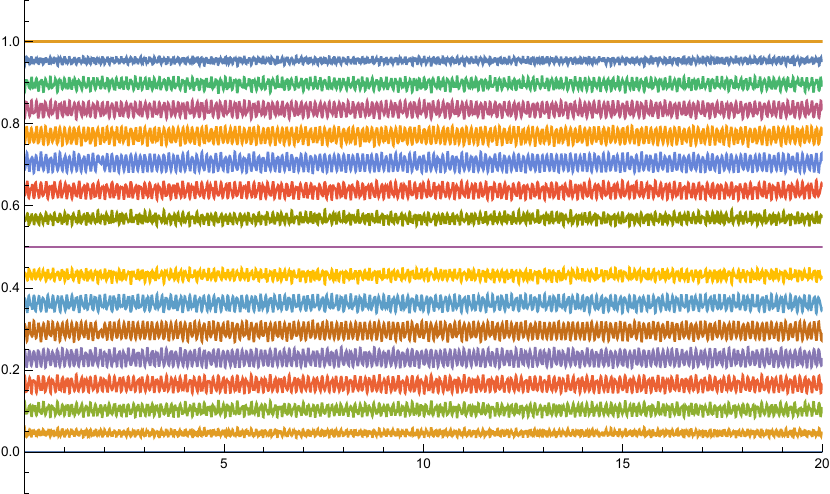}
\end{center}
\end{figure}

As can be seen, the charges oscillate around a mean position independent of $t$. In particular, the charges do not approach a state of equilibrium in which state it would not move anymore. The central loads are further away than the loads located at the ends. This is consistent with a remark by Griffiths and Li. Note that the oscillations of individual charges are getting smaller when the number of charges on the needle increases. This point is very important, as we will see at the end of the paper.

\section*{2. Equilibrium}

\subsection*{2.1 --- The system of equations to solve}

Consider $n>2$ identical charges placed on the segment $[0,1]$ whose respective positions are $x_1<x_2< \cdots <x_n$. Each charge experiences the force exerted on it by the other charges. The leftmost charge must therefore be at $0$, and the rightmost charge must be at $1$. We therefore have $x_1=0$ and $x_n=1$. If $i \in \llbracket 2,n-1 \rrbracket$ (not for $i\in \{ 0,n \}$), the force generated on the charge in $x_i$ by the charges placed on the left must be compensated by the one generated by the charges placed on the right. We must therefore have $\sum\limits_{j=1}^{i-1} \frac{1}{\left(x_i-x_j\right)^2} = \sum\limits_{j=i+1}^n \frac{1}{\left(x_i-x_j\right)^2}$. If we know how to solve this system of equations, we will know the equilibrium position of the charges. For all $n \geqslant 3$, the system $\left( E_n \right)$ writes 

\[ \left( E_n \right) \quad \begin{cases}
x_1 = 0&\\
\sum\limits_{j=1}^{i-1} \frac{1}{(x_i - x_j)^2} = \sum\limits_{j=i+1}^n \frac{1}{(x_i - x_j)^2}&\text{if $2 \leqslant i \leqslant n-1$}\\
x_n = 1&
\end{cases}. \]

So let's study the system a little more closely for small values of $n$.

\textbullet\ If $n=3$, no need for calculation : $\left( x_1, x_2, x_3 \right) = \left( 0, \frac{1}{2}, 1 \right)$.

\textbullet\ If $n=4$, by symmetry, we must have $\left( x_1, x_2, x_3, x_4 \right)=(0, \alpha , 1-\alpha , 1)$, with $\alpha \in  ]0, \left.\frac{1}{2}\right[$. The system is reduced to a single equation,
\[ \frac{1}{\alpha^2} = \frac{1}{(1- \alpha )^2} + \frac{1}{((1- \alpha ) - \alpha )^2} = \frac{1}{(1- \alpha )^2} + \frac{1}{(1-2 \alpha )^2} . \]

Solving this equation uses the roots of the polynomial $X^4+6X^3-11X^2+6X-1$. Numerically, we obtain $\alpha \simeq 0.319$. We could solve the equation explicitly.

But as soon as $n>4$, we obtain polynomial systems of high degree, and we know that there is no exact closed formula for this resolution.

\textbullet\ If \(n=6\), by symmetry, we must have 
\[ \left( x_1, x_2, x_3, x_4, x_5, x_6 \right) = (0, \alpha , \beta ,1-\beta , 1-\alpha , 1), \]
with $0<\alpha <\beta <\frac{1}{2}$. The system is \\

\noindent \resizebox{\linewidth}{!}{$\left\{ \begin{array}{lcl}
\frac{1}{\alpha ^2} &= &\frac{1}{(\beta -\alpha )^2}+\frac{1}{(1-\beta -\alpha )^2}+\frac{1}{(1-2\alpha )^2}+\frac{1}{(1-\alpha )^2} \\
\frac{1}{\beta ^2}+\frac{1}{(\beta -\alpha )^2} &= &\frac{1}{(1-2\beta )^2}+\frac{1}{(1-\alpha -\beta )^2}+\frac{1}{(1-\beta )^2} 
\end{array} \right..$}\\

When we solve this system, we see that $\alpha$ and $\beta$ are two of the roots of a polynomial of degree 32 (not reproduced here).

More generally, as soon as $n > 4$, we obtain (after increasingly long computations) a polynomial of high degree, of which we must find $n$ roots, and we know that there is no exact closed formula for this resolution.

For this reason, we must move towards digital methods.

\subsection*{2.2 --- Proof of the existence of the state of equilibrium and first method of numerical determination of this equilibrium}

In this subsection, we are going to prove the existence of an equilibrium point. As $\sum\limits_{j=1}^{i-1} \frac{1}{\left(x_i-x_j\right)^2}-\sum\limits_{j=i+1}^n \frac{1}{\left(x_i-x_j\right)^2}$ is unbounded, it would be hard to do by using the system $\left( E_n \right)$ of equations. For that reason, we are going to use another mathematical method. We are going to construct a function $\varphi$, purely mathematical, that is without any physical meaning, which has nice mathematical properties: it is continuous (and also differentiable) on a domain which includes the domain in which are the positions $\left( x_1, x_2, \ldots , x_n \right)$, and most important : a point $\left( x_1, x_2, \ldots , x_n \right)$ is a fixed point of $\varphi$ if and only if it is an equilibrium point, that is a solution of the system studied in the former subsection, $\left( E_n \right)$.

Let $n$ be an integer, $n \geqslant 4$. The part of $\mathbb{R}^n$ in which we look for the $n$-tuple $\left( x_1, x_2, \ldots , x_n \right)$ is\\

\noindent \resizebox{\linewidth}{!}{$S_n = \Set{(a_1 , a_2 , \ldots , a_n) \in \mathbb{R}^n | 0 \leqslant a_1 < a_2 < \cdots < a_n \leqslant 1} .$}\\

It is a bounded simplex. To be able to apply the fixed point theorems, we considered its closure \\

\noindent \resizebox{\linewidth}{!}{$\overline{S_n} = \Set{(a_1 , a_2 , \ldots , a_n) \in \mathbb{R}^n | 0 \leqslant a_1 \leqslant a_2 \leqslant \cdots \leqslant a_n \leqslant 1} .$}\\

The disadvantage of this set is that when $x_i=x_j$, the term $\frac{1}{\left(x_i-x_j\right)^2}$ is not defined. To get around this difficulty, we used the function $x \mapsto \exp \left( \frac{-1}{x^2} \right)$ which extends to $0$, with a $\mathcal{C}^{\infty}$ extension, whose successive derivatives at $0$ are all zero. Below, the extended function is again denoted in the same way.

Let $\varphi$ the function from $\overline{S_n}$ to $\overline{S_n}$ defined as follows. For all $\left( x_1, x_2, \ldots , x_n \right) \in \overline{S_n}$, we have $\varphi \left( x_1, x_2, \ldots , x_n \right) = \left( y_1, y_2, \ldots , y_n \right)$ with

\textbullet\ $y_1=0$.

\textbullet\ $y_n=1$.

\textbullet\ For all $i\in \llbracket 2,n-1 \rrbracket$, 

\noindent\resizebox{\linewidth}{!}{$y_i = \begin{cases}
x_i&\text{if $x_{i-1} = x_i = x_{i+1}$}\\
x_i - \frac{1}{3}(x_i - x_{i-1})&\text{if $x_{i-1} \neq x_i = x_{i+1}$}\\
x_i + \frac{1}{3} (x_{i+1} - x_i)&\text{if $x_{i-1} = x_i \neq x_{i+1}$}\\
\begin{array}{l}
x_i + \frac{1}{3} \min(|x_i - x_{i-1}| , |x_{i+1} - x_i|) \\
\times \sgn \left( \sum\limits_{j=1}^{i-1} \frac{1}{(x_i - x_j)^2} - \sum\limits_{j=i+1}^n \frac{1}{(x_i - x_j)^2} \right) \\
\times \exp \left( \frac{-1}{\left( \sum\limits_{j=1}^{i-1} \frac{1}{(x_i - x_j)^2} - \sum\limits_{j=i+1}^n \frac{1}{(x_i - x_j)^2} \right)^2} \right)
\end{array}&\text{otherwise}
\end{cases}.$}\\

The function $\varphi$ is continuous, and a point $\left( x_1, x_2, \ldots , x_n \right) \in \overline{S_n}$ is an equilibrium point if and only if it is a fixed point of $\varphi$. As $\overline{S_n}$ is homeomorphic to the unit ball of $\mathbb{R}^n$, Brouwer's theorem asserts that $\varphi$ has a fixed point, therefore that \textbf{the system of charges placed on the needle has a equilibrium point}.

For all $\left(x_1, x_2, \ldots , x_n \right) \in \overline{S_n}$, the sequence $\left( \varphi^p \left( x_1, x_2, \ldots , x_n \right) \right)$ of the iterates of $\varphi$ converges to a fixed point of $\varphi$. Using these iterations, we determined numerical approximations of equilibrium points for a certain number of values of $n$. Note however that as the successive derivatives of $x \mapsto \exp \left( \frac{-1}{x^2} \right)$ at $0$ are all zero, the iterates converge towards the fixed point very slowly. For this reason, in order to approximate the fixed point, we used other algorithms.

\subsection*{2.3 --- Proof of the uniqueness of the state of equilibrium}

Let $n$ be an integer, $n \geqslant 4$. Considere $n$ charges at equilibrium on the needle. Then, $x_1=0$, and $x_n=1$. So we concentrate on the $n-2$ remaining charges. The part of $\mathbb{R}^{n-2}$ in which we look for the $n-2$-tuple $\left(x_2, x_3, \ldots , x_{n-1} \right)$ is 
\resizebox{\linewidth}{!}{$U_{n-2} = \Set{(a_2 , \ldots , a_{n-1}) \in \mathbb{R}^n | 0 < a_2 <a_3 < \cdots < a_{n-1} < 1}.$}

The set $U_{n-2}$ is an open, bounded and convex subset of $\mathbb{R}^{n-2}$. For any $\left( a_2, \ldots , a_{n-1} \right)\in U_{n-2}$, we set $a_1=0$ and $a_n=1$. Let 
\[ \omega : \left\lbrace
  \begin{array}{rcl}
    U_{n-2} & \to & \mathbb{R} \\
    (a_2 , \ldots , a_{n-1}) & \mapsto & \sum\limits_{j \in \llbracket 1,n-1 \rrbracket \atop {k \in \llbracket 2,n \rrbracket \atop j<k}} \frac{1}{a_k - a_j} \\
  \end{array}
\right.. \]

Then, $\omega$ is continuous and differentiable. Let us show that it is a strictly convex function. Let $A= \left( a_2, \ldots , a_{n-1} \right)$ and $B=\left( b_2, \ldots, b_{n-1} \right) \in U_{n-2}$. Let 
\[ f: \left\lbrace
  \begin{array}{rcl}
    [0,1] & \to & \mathbb{R} \\
    t & \mapsto & \omega (t A + (1-t) B) \\
  \end{array}
\right.. \]

Then 
\begin{align*}
f(t) &= \sum\limits_{j \in \llbracket 1,n-1 \rrbracket \atop {k \in \llbracket 2,n \rrbracket \atop j<k}} \frac{1}{t a_k+(1-t) b_k - t a_j-(1-t)b_j} \\
&= \sum\limits_{j \in \llbracket 1,n-1 \rrbracket \atop {k \in \llbracket 2,n \rrbracket \atop j<k}} \frac{1}{t \left(a_k-b_k-a_j+b_j\right)+\left( b_k - b_j\right)}
\end{align*}
\[ f'(t) = - \sum\limits_{j \in \llbracket 1,n-1 \rrbracket \atop {k \in \llbracket 2,n \rrbracket \atop j<k}} \frac{a_k-b_k-a_j+b_j}{\left( t \left( a_k-b_k-a_j+b_j \right)+ \left( b_k - b_j \right) \right)^2} \]
\[ f''(t) = \sum\limits_{j \in \llbracket 1,n-1 \rrbracket \atop {k \in \llbracket 2,n \rrbracket \atop j<k}} \frac{2 \left( a_k-b_k-a_j+b_j \right)^2}{\left( t \left( a_k-b_k-a_j+b_j \right)+\left( b_k - b_j \right) \right)^3}. \]

Since the sum is made with the asumption $j<k$, each term of the last sum is positive, which shows that $\omega$ is strictly convex. So, $\omega$ is a differentiable function, strictly convex, having an infinite limit on the boundary of the open convex bounded domain $U_{n-2}$. Therefore, this function has an unique minimum, which is a critical point, that is a point where the gradient of $\omega$ is null.

In order to simplify, we note $\left( e_2, \ldots , e_{n-2} \right)$ the canonical basis of $\mathbb{R}^{n-2}$, and we use the common abbreviation $\partial _k$ for $\partial _{e_k}$, the partial derivative in the direction of $e_k$. With this notation, we get 
\[ \partial _k\omega \left( a_2, \ldots , a_{n-1} \right) = \sum\limits_{j=k+1}^n \frac{1}{(a_k - a_j)^2} - \sum\limits_{j=1}^{k-1} \frac{1}{(a_k - a_j)^2} . \]

The gradient of $\omega$ at a point $A=\left(a_2, \ldots , a_{n-1}\right)\in U_{n-2}$ is null if and only if $A$ is an equilibrium point of $\left(a_1, a_2, \ldots , a_{n-1}, a_n \right)$. 

This shows the uniqueness of the equilibrium point, and provides a second proof of the existence.

\subsection*{2.4 --- Usefulness of the Two Proofs of the Existence of \(\text{Equilibrium}\)}

As mentioned above, the iterative method using $\varphi$ defined in Section 2.2 converges very slowly. The proof in Section 2.3 provides access to gradient-descent algorithms. Unfortunately, these algorithms can only be used if the starting point is very close to the desired solution. Otherwise, the points found do not belong to the domain $S_n$. For these reasons, for the numerical search for equilibrium points, we first used an ad-hoc algorithm based on the $\varphi$ function, followed, to improve accuracy, by a gradient-descent algorithm.

Note the following points regarding the computational complexity: When the number of charges is doubled, for the equilibrium point calculation,

\textbullet\ The number of points to be computed is doubled;

\textbullet\ The computations of each point involves twice as many points (the number of charges affecting a given charge is doubled);

\textbullet\ The algorithm's step size must be divided by 2 so that the approximations found belong to $S_n$;

\textbullet\ The precision of the computations (the number of digits with which the computations are performed) must be increased to control error propagation during the computations.

\section*{3. Two differential methods for numerically assessing the steady state}

\subsection*{3.1 --- Differential system of charges' move}

We saw above that charges oscillate indefinitely; the curves of the trajectories of the differential system studied show that these oscillations take place around a point of equilibrium. Suppose we have calculated an approximate solution 
\[ \sigma: \left\lbrace
  \begin{array}{rcl}
    [0,a] & \to & S_n \\
    t & \mapsto & X(t) = \begin{pmatrix}
    x_1 (t) \\
    x_2 (t) \\
    \vdots \\
    x_n (t)
\end{pmatrix}     \\
  \end{array}
\right. \]

\noindent of the differential system (\ref{systemone}) given above, with $a>1$. Then $\frac{1}{a-1} \int_1^a X(t) \mathrm{d}t$ provides an approximation of the equilibrium. Taking the integral from $1$ erases part of the effect of the initial position. Note that using this differential equation requires a very large amount of calculations. Here is another differential equation also giving a means of calculating the equilibrium point, but requiring fewer calculations.

\subsection*{3.2 --- Another differential system}

The differential system that we will consider here is directly inspired by the differential system (\ref{systemone}), but it is independent of it. Here is this autonomous system.

Let $\left(a_1=0, a_2, \ldots ,a_n=1 \right) \in S_n$.

\begin{equation}
\resizebox{0.9\linewidth}{!}{$\begin{cases}
x_1 (t) =0&\\
x_n (t) = 1&\\
x_i' (t) = \sum\limits_{j=1}^{i-1} \frac{1}{(x_i (t) - x_j (t))^2} - \sum\limits_{j=i+1}^n \frac{1}{(x_i (t) - x_j (t))^2}&\text{for all $i \in \llbracket 2,n-1 \rrbracket$}\\
x_i (0) = a_i&\text{for all $i \in \llbracket 1,n \rrbracket$}
\end{cases}.
\label{systemtwo}$}
\end{equation}

Let $t \mapsto X(t)$ be the solution of the system, unique according to Cauchy's theorem. Still according to this theorem, $X$ must leave any compact in which it enters. Since the edge of $S_n$ is formed by vectors of $\mathbb{R}^n$ having at least two equal coordinates, and since $\frac{1}{\left(x_i-x_j\right)^2}$ tends to $+ \infty$, when $x_i-x_j$ tends towards $0$, $X$ is defined on $[0,\infty [$. Since $X([0,\infty[)$ is included in the compact $\overline{S_n}$, so is $\overline{X([0,\infty [)}$ which is compact, connected and without isolated points. If $X([0,\infty [)=\overline{X([0,\infty [)}$, the solution is asymptotically periodic. It is impossible for the minimum period not to be zero because, by system definition, for all $i\in \llbracket 2,n-1 \rrbracket$, $\left| \sum\limits_{j=1}^{i-1} \frac{1}{\left(x_i(t) - x_j(t)\right)^2}-\sum\limits_{j=i+1}^n \frac{1}{\left(x_i(t) - x_j(t)\right)^2} \right|$ is decreasing. This same reason prevents $\overline{X([0, \infty[)} \setminus X([0, \infty[)$ from containing two distinct points. There therefore remains only one possibility: When $t \longrightarrow \infty$, $X(t)$ tends towards an equilibrium point. This provides another proof of the existence of that equilibrium.

Numerical resolutions on small values of $n$ show that $X(t)$ tends towards the same equilibrium point as the other algorithms, which is coherent with the uniqueness proved above. Note that in order to get a numerical solution, one has to be very carefull : a very small step must be taken, because otherwise the proposed solution comes out of $S_n$. For the approximate computation of the solutions, the complexity is high. For example, if we double the number of points, each evaluation of a $x_i'(t)$ requires twice as many calculations. There are also twice as many points to calculate, and the calculation step must be divided by $2$, while the precision of the calculations must be increased. In short, a multiplication by $2$ of the number of points multiplies the calculation time by more than $2^3=8$.

For the computations, we took as initial condition $\left( a_1=0, a_2, \ldots , a_n=1 \right)\in S_n$, with $a_i=\frac{i-1}{n}$. Here are the trajectories obtained for the first values of $n=2^p+1$.

$n=2^2+1=5$

\begin{figure}[H]
\begin{center}
\includegraphics[scale=0.75]{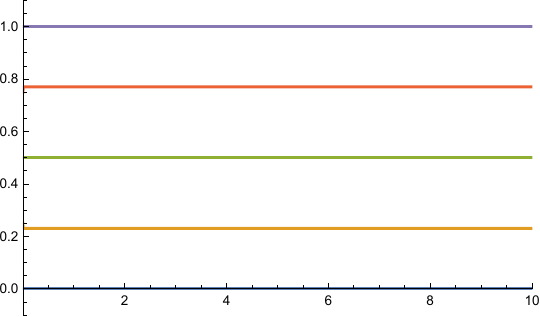}
\end{center}
\end{figure}

$n=2^3+1=9$

\begin{figure}[H]
\begin{center}
\includegraphics[scale=0.75]{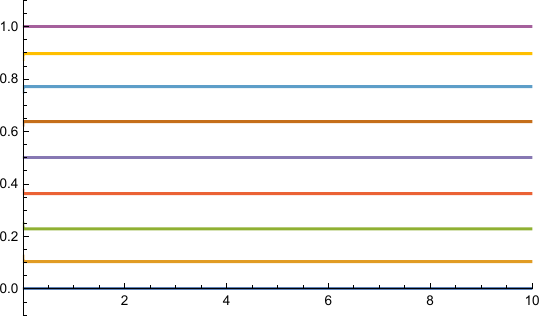}
\end{center}
\end{figure}

$n=2^4+1=17$

\begin{figure}[H]
\begin{center}
\includegraphics[scale=0.75]{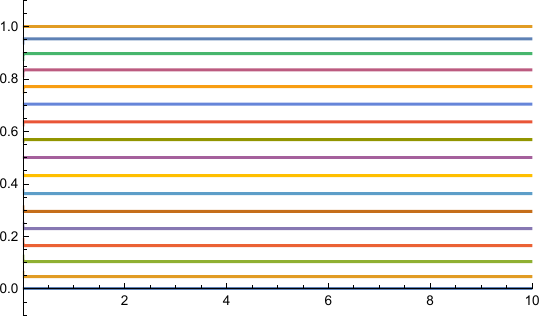}
\end{center}
\end{figure}

$n=2^5+1=33$

\begin{figure}[H]
\begin{center}
\includegraphics[scale=0.75]{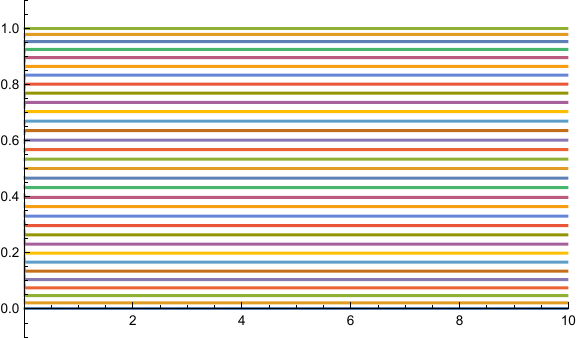}
\end{center}
\end{figure}

As seen in these graphs, the trajectories appear to be straight lines. This is because the convergence towards the equilibrium point is fast. If we look at the far left of the trajectories, we see that in fact these trajectories are not straight lines. We also observe that the equilibrium distribution concentrates a little more charges on the edges of the needle than in the center. Once again this confirms a remark made by Griffiths and Li.

\subsubsection*{Remark}

The mere observation of the formula used in system (\ref{systemtwo}) could suggest that system (\ref{systemtwo}) is the one which corresponds to the derivative of a solution
\(X\) of (\ref{systemone}). It is not so. The domain of definition of a differential system is essential. Let us clarify this point.

Let 
\[ \theta: \left\lbrace
  \begin{array}{rcl}
    S_n & \to & \mathbb{R}^n \\
    (x_i)_{1 \leqslant i \leqslant n} & \mapsto & (y_i)_{1 \leqslant i \leqslant n}     \\
  \end{array}
\right. \]

\noindent with \\

\noindent \resizebox{\linewidth}{!}{$y_i = \begin{cases}
0&\text{if $i=1$}\\
\sum\limits_{j=1}^{i-1} \frac{1}{(x_i(t) - x_j(t))^2} - \sum\limits_{j=i+1}^n \frac{1}{(x_i(t) - x_j(t))^2}&\text{if $i \in \llbracket 2,n-1 \rrbracket$}\\
1&\text{if $i=n$}
\end{cases}.$}

Let us denote $X = \begin{pmatrix}
x_1 \\
x_2 \\
\vdots \\
x_n
\end{pmatrix}$ and $Y = \begin{pmatrix}
X' \\
X
\end{pmatrix}$. Then, in Cauchy form, system (\ref{systemone}) is written $Y'=\eta (Y)$, with $\eta \left( \begin{pmatrix}
U \\
V 
\end{pmatrix} \right) = \begin{pmatrix}
\theta (V) \\
U 
\end{pmatrix}$. The domain of $\eta$ is $\mathbb{R}^n \times S_n$. In particular, $X'$ can for example take the value $0$ (the vector), which does not belong to $S_n$.

The Cauchy form of system (\ref{systemtwo}) is $X'=\theta (X)$, with the domain $S_n$, which does not contain $0$.

Thus, if $X$ is a solution of system (\ref{systemone}), $X'$ is not a solution of (\ref{systemtwo}).

\section*{4. Adding a new charge}

\subsection*{4.1. First order move}

We start from $n$ charges at equilibrium. The positions of the charges are $x_1, \ldots , x_n$. We add a charge between charges numbers $i$ and $i+1$, with $i<n/2$. Due to the introduction of this new charge, the charges from $1$ to $i$ move to the left, and the charges from $i+1$ to $n$ move to the right. Their numbering also changes. The charge $i+1$ takes the number $i+2$, ... and $n$ becomes $n+1$. We note $y_1, \ldots , y_{n+1}$ the equilibrium positions of the $n+1$ charges.

Charges from $1$ to $i$ go from $x_k$ to $y_k$ (with $k \in \llbracket 1,i \rrbracket$). For any $k\in \llbracket i+1,n \rrbracket$, the charge of number $k$ at start goes from $x_k$ to $y_{k+1}$. The following graphics shows these moves : $y_k-x_k$ if $k\in \llbracket 1,i \rrbracket$, $y_{k+1}-x_k$ if $k\in \llbracket i+1,n \rrbracket$, for the peculiar case $n=513$, $i=194$ (the point for $194$ has no meaning).

\begin{figure}[H]
\begin{center}
\includegraphics[scale=0.9]{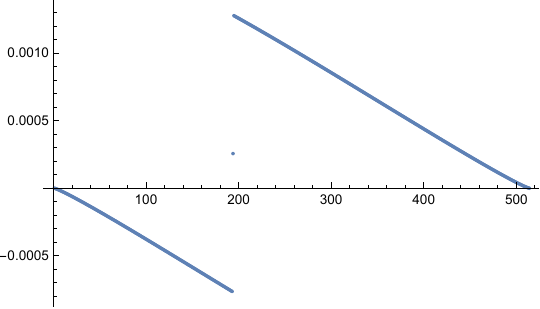}
\end{center}
\end{figure}

As this graphics illustrates, numerical experiments studying the displacements of the charges show that these moves are to the first order piecewise linear, on $\llbracket 1,i \rrbracket$ and on $\llbracket i+2,n+1 \rrbracket$, with a slope of order around $\frac{-1}{n^2}$ (as a function of $k$). Charges from $1$ to $i$ go from $x_k$ to $y_k=x_k-\frac{\alpha}{n^2} (k-1)$ (with $k\in \llbracket 1,i \rrbracket$). For any $k\in \llbracket i+1,n \rrbracket$, the charge of number $k$ at start goes from $x_k$ to $y_{k+1}$, with $y_{k+1}=x_k+\frac{\alpha}{n^2} (n-k)$.

The equilibrium positions satisfy a symmetry condition.
\[ \begin{array}{c}
\text{for all } k \in \llbracket 1, n \rrbracket , x_k = 1- x_{n+1-k} \\
\text{for all } q \in \llbracket 1, n+1 \rrbracket , y_q = 1-y_{n+2-q} 
\end{array} \]

In addition, the equilibrium position of the $n+1$ charges is independent of the location to which a charge has been added.

Below we will explore the consequences of these properties for the calculation of $y_q$, assuming that the $y_i$-s admit an expansion of order at least $2$. We will successively see $q \in \llbracket 1, i \rrbracket \cup \llbracket n+2-i , n+1 \rrbracket$, $q=i+1$, $q \in \llbracket i+2 , n-i \rrbracket$, then we will summarize the results.\\

\textbullet\ $y_q$, with $q \in \llbracket 1, i \rrbracket \cup \llbracket n+2-i , n+1 \rrbracket$\\

We of course have $y_1=0$ and $y_{n+1}=1$. 

To begin with, suppose that for all $k\in \llbracket i+1,n \rrbracket$, the load of number $k$ at the start goes from $x_k$ to $y_{k+1}$, with $y_{k+1}=x_k+\frac{\beta}{n^2} (n-k)$. We will see why $\alpha =\beta$.

Let's start with $y_2$. We have $y_2=x_2-\frac{\alpha}{n^2}$. Due to the symmetry of the $x_k$-s, we also have $y_2=\left(1-x_{n-1}\right)-\frac{\alpha}{n^2}$. Due to the symmetry of $y_k$-s, we have $y_2=1-y_n=1-\left( x_{n-1}+\frac{\beta}{n^2} \right)$. As a result, $1-x_{n-1}-\frac{\alpha}{n^2}=1-x_{n-1}-\frac{\beta}{n^2}$, hence $\alpha =\beta$. 

More generally, if $k\in \llbracket n+2-i, n+1 \rrbracket$,

To first order, we therefore have 
\[ \text{for all } k \in \llbracket 1,i \rrbracket , y_k=x_k-\frac{\alpha}{n^2}(k-1) \]

\noindent and 
\[ \text{for all } k\in \llbracket n+2-i, n+1 \rrbracket , y_k=x_{k-1}-\frac{\alpha}{n^2} (n+1-k). \]

\textbullet\ $y_{i+1}$\\

This is the new charge. We do not directly deduce the position of this charge from the position of an old charge. We can nevertheless evaluate this position by symmetry.

We observe that $y_{i+1}$ satisfies the same equation as $y_k$, $k \in \llbracket 1,i \rrbracket$. This is in agreement with the observation that the equilibrium position of the $n+1$ charges does not depend on the location at which the new charge was added.\\

\textbullet\ $y_q$, with $q \in \llbracket i+2 , n-i \rrbracket$\\

Let's start with $y_{i+2}$. We have 
\begin{align*}
y_{i+2} &=x_{i+1}+\frac{\alpha}{n^2}(n-i-1) \\
&=1-y_{n+2-(i+2)} \\
&=1-y_{n-i} \\
&=1-\left( x_{n-i-1}+\frac{\alpha}{n^2}(i+1) \right) \\
&=1-x_{n-i-1}-\frac{\alpha}{n^2}(i+1) \\
&=x_{i+2}-\frac{\alpha}{n^2}(i+1)
\end{align*}

We must therefore have

\[ x_{i+1}+\frac{\alpha}{n^2} (n-i-1)=x_{i+2}-\frac{\alpha}{n^2} (i+1) \]

\noindent that is

\[ \frac{\alpha}{n^2} (n-i-1+i+1)=x_{i+2}-x_{i+1} \]

\noindent or

\[ \alpha = n \left( x_{i+2}-x_{i+1} \right) . \]

Thus, the coefficient $\alpha$ actually depends on the interval considered. The first $n$ charges split the interval into $n-1$ pieces, and on average, consecutive charges are at a distance of $1/(n-1)$. We therefore have approximately $\alpha =\frac{n}{n-1}$, which explains that in numerical experiments, we have the feeling of a slope of $\frac{1}{n^2}$, corresponding to $\alpha =1$. Let's report in the expression $y_{i+2}$. We obtain 
\begin{align*}
y_{i+2} &=x_{i+1}+\frac{\alpha}{n^2}(n-i-1) \\
&=x_{i+1}+\frac{n-i-1}{n^2} n \left( x_{i+2}-x_{i+1} \right) \\
&=x_{i+1}+x_{i+2}-\frac{i+1}{n} x_{i+2}-x_{i+1}+\frac{i+1}{n} x_{i+1} \\
&=x_{i+2}-\frac{i+1}{n} \left( x_{i+2}-x_{i+1} \right) .
\end{align*}

More generally, for any $q\in \llbracket i+2, n-i \rrbracket$, 
\begin{align*}
y_q &= x_{q-1}+\frac{\alpha}{n^2} (n+1-q) \\
&=1-y_{n+2-q} \\
&=1- \left( x_{n+1-q}+\frac{\alpha}{n^2} (q-1) \right) \\
&=1- \left( 1-x_q+\frac{\alpha}{n^2} (q-1) \right) \\
&=x_q-\frac{\alpha}{n^2} (q-1)
\end{align*}

\noindent and therefore 
\[ x_{q-1}+\frac{\alpha}{n^2} (n+1-q)=x_q-\frac{\alpha}{n^2} (q-1) \]

\noindent then 
\[ \frac{\alpha}{n^2}(n+1-q+q-1)=x_q-x_{q-1} \]

\noindent that is 
\[ \alpha = n \left( x_q-x_{q-1} \right) \]

\noindent hence 
\begin{align*}
y_q &= x_q- \frac{\alpha}{n^2} (q-1) \\
&= x_q- \frac{q-1}{n} \left( x_q-x_{q-1} \right) .
\end{align*}

The formulas that we obtain ultimately do not depend on the location where we add the new charge, which is consistent with the remark made above.

We can group these formulas into one: 

\begin{prop}
If $x_1 = 0 < x_2 < \cdots < x_{n-1} < x_n = 1$ are the positions of $n$ charges at equilibrium, then, if $n+1$ charges at $y_1 < y-2 < \cdots < y_{n+1}$ are at equilibrium, 
\noindent \resizebox{\linewidth}{!}{$\begin{cases}
y_1 = 0&\\
y_k = \frac{(k-1)}{n} x_{k-1} + \frac{n+1-k}{n} x_k + O \left( \frac{1}{n^3} \right) ,&\text{for all $k \in \llbracket 2,n \rrbracket$}\\
y_{n+1} = 1&
\end{cases} .$}
\end{prop}

We will now test the validity of these conjectures and see some consequences.

\subsection*{4.2. Computation for $y_2$}

Applied to $k=2$, the former formula gives $y_2=\frac{n-1}{n} x_2$. If we note $X_{n,2}$, the position of the second charge of a system of $n$ charges at equilibrium, we get successively $X_{n+1,2}=\frac{n-1}{n} X_{n,2}$, $X_{n+2,2}=\frac{n}{n+1} X_{n+1,2}=\frac{n}{n+1} \times \frac{n-1}{n} X_{n,2}=\frac{n-1}{n+1} X_{n,2}$, ..., $X_{2n-1,2}=\frac{n-1}{2n-2} X_{n,2}=\frac{1}{2} X_{n,2}$. We tested the formula and compared to the true computed value for $n\in \{5,9,17,33,65,129,257,513,1025,2049\}$.
Here is the result: 
\[ \begin{array}{|c|c|}
\hline
 n & \frac{X_{n,2}}{X_{2 n-1,2}} \\
\hline
 5 & 2.226 \\
\hline
 9 & 2.229 \\
\hline
 17 & 2.212 \\
\hline
 33 & 2.189 \\
\hline
 65 & 2.166 \\
\hline
 129 & 2.14627 \\
\hline
 257 & 2.12952 \\
\hline
 513 & 2.1156 \\
\hline
 1025 & 2.10893 \\
\hline
\end{array} \]

The greater is $n$, the better is the approximation, which is coherent for an order $1$ approximation.

\subsection*{4.3. Convexity}

The formula $y_k=\frac{(k-1)}{n} x_{k-1}+\frac{n+1-k}{n} x_k + O \left( \frac{1}{n^3} \right)$ for $k\in \llbracket 2,n \rrbracket$ means that $y_k$ is between $x_{k-1}$ and $x_k$. This can be proved directly: if we think that the sequence $\left(y_1, \ldots , y_{n+1} \right)$ has been obtained from $\left(x_1, \ldots , x_n \right)$ by adding a charge between $x_i$ and $x_{i+1}$ with $i+1<k$, we can think that $y_k$ is the new position of the charge $x_{k-1}$, which was pushed to the right by the new charge. So, $y_k>x_{k-1}$. If on the contrary we think that that the sequence $\left( y_1, \ldots , y_{n+1} \right)$ has been obtained from $\left( x_1, \ldots , x_n \right)$ by adding a charge between $x_i$ and $x_{i+1}$ with $i>k$, we can think that $y_k$ is the new position of the charge $x_k$, which was pushed to the left by the new charge. So, $y_k < x_k$.

\subsection*{4.4. Length of the intervals between consecutive charges}

Experiments show that the length of the interval between two consecutive charges is maximal at the center of $[0,1]$, and the length is decreasing from the center to the ends of $[0,1]$. Assuming that the former approximation is acute, let us show that this monotony persists when adding a new charge. That will prove that this is true whatever the number of charges is.

Consider $n$ charges at equilibrium, at positions $x_1=0$, $x_2$, ..., $x_n=1$. Add a new charge. Let $y_i$, $1 \leqslant i \leqslant n+1$ the positions of the charges at equilibrium. We consider charges in $\left[ \frac{1}{2},1 \right]$. Assume $k>\frac{n}{2}$, $x_{k-1}<y_k<x_k<y_{k+1}<x_{k+1}<y_{k+2}<x_{k+2}$. We suppose also 
\[ x_k-x_{k-1}>x_{k+1}-x_k>x_{k+2}-x_{k+1}. \]

We want to prove that $y_{k+1}-y_k>y_{k+2}-y_{k+1}$. According to the approximation given above, we have 
\[ \left\{ \begin{array}{lcl}
y_k &= &\frac{k-1}{n} x_{k-1}+\frac{n+1-k}{n} x_k \\[0.3em]
y_{k+1} &= &\frac{k}{n} x_k+\frac{n-k}{n} x_{k+1} \\[0.3em]
y_{k+2} &= &\frac{k+1}{n} x_{k+1}+\frac{n-k-1}{n} x_{k+2}
\end{array} \right.. \]

Therefore 
\[ \left\{ \begin{array}{lcl}
y_{k+2}-y_{k+1} &= &\frac{n-k-1}{n} x_{k+2}+\frac{2k+1-n}{n} x_{k+1}-\frac{k}{n} x_k \\[0.3em]
y_{k+1}-y_k &= &\frac{n-k}{n} x_{k+1}+\frac{2k-1-n}{n} x_k-\frac{k-1}{n} x_{k-1}
\end{array} \right. \]

\noindent and 
\[ \left\{ \begin{array}{lcl}
y_{k+2}-y_{k+1} &= &\frac{n-k-1}{n} \left( x_{k+2}-x_{k+1} \right)+\frac{k}{n} \left( x_{k+1}-x_k \right) \\[0.3em]
y_{k+1}-y_k &= &\frac{n-k}{n} \left( x_{k+1}-x_k \right) + \frac{k-1}{n} \left( x_k-x_{k-1} \right)
\end{array} \right.. \]

Then 
\[ \left\{ \begin{array}{lcl}
y_{k+2}-y_{k+1} &= &\frac{n-k}{n} \left( x_{k+2}-x_{k+1} \right) + \frac{k-1}{n} \left( x_{k+1}-x_k \right)\\[0.15em]
& &+ \frac{1}{n} \left( \left( x_{k+1}-x_k \right) - \left( x_{k+2}-x_{k+1} \right) \right)
\\[0.3em]
y_{k+1}-y_k &= &\frac{n-k}{n} \left( x_{k+1}-x_k \right) + \frac{k-1}{n} \left( x_k-x_{k-1} \right)
\end{array} \right.. \]

It follows that \\

\noindent \resizebox{\linewidth}{!}{$\begin{array}{lcl}
\left( y_{k+1}-y_k \right)- \left( y_{k+2}-y_{k+1} \right) &= &\frac{n-k}{n} \left( \left( x_{k+1}-x_k \right) - \left( x_{k+2}-x_{k+1} \right) \right) \\[0.15em]
& &+\frac{k-1}{n} \left( \left( x_k-x_{k-1} \right) - \left( x_{k+1}-x_k \right) \right) \\[0.15em]
& &-\frac{1}{n} \left( \left( x_{k+1}-x_k \right) - \left( x_{k+2}-x_{k+1} \right) \right) \\[0.3em]
&= &\frac{k-1}{n} \left( \left( x_k-x_{k-1} \right) - \left( x_{k+1}-x_k \right) \right) \\[0.15em]
& &+\frac{n-k-1}{n} \left( \left( x_{k+1}-x_k \right) - \left( x_{k+2} - x_{k+1} \right) \right) \\[0.3em]
&> &0.
\end{array}$}

\section*{5. Dyadic density approximation}

To approach the continuous charge density, it is necessary to study the distribution of elementary charges for an increasing number of charges. We used divisions of $[0,1]$ into $2^p$ intervals. A first advantage of this type of division is that once we have a point having a proportion of $\frac{q}{2^s}$ points to its left and $\frac{2^s-q}{2^s}$ points to its right, there has such a point in all the following cuts. As we will see below, this makes it possible to obtain the distribution function associated with the density we look for. Let's explain with two examples. Let us denote $\left( a_1, a_2, a_3, a_4, a_5 \right)$ the equilibrium point for $5=2^2+1$ charges, $\left( b_1, \ldots , b_9 \right)$ the one for $9=2^3+1$ loads, and $\left( c_1, \ldots , c_{17} \right)$ the one for $17=2^4+1$. Then $a_2$ has three times as many points to its right as to its left. The same holds for $b_3$ and $c_5$. In the sequence $\left( b_1, \ldots , b_9\right)$, $b_6$ has $\frac{5}{8}$ of the other points to its left, which is found in the series $\left( c_1, \ldots , c_{17} \right)$ with $c_{11}$. More generally, take any dyadic number $\gamma =\frac{p}{2^n}$, with $p$ an odd number $<2^n$. Then, for any $q \geqslant n$ there exists a single $x_{m_{q,\gamma}}$, with $1 \leqslant m_{q,\gamma} \leqslant 2^q$ such that in the sequence $x_1<x_2< \cdots <x_{m_{q,\gamma}}< \cdots <x_{2^q+1}$, there are exactely $\gamma \times 2^q$ numbers $x_i$ which are strictly smaller than $x_{m_{q,\gamma}}$. In other words, $m_{q,\gamma}-1=\gamma \times 2^q$.

The approximate values of the equilibrium points were calculated for $5$, $9$, $17$, ..., $4097$ charges.

Here is a graph showing the charges distributions for these values. In red, $\gamma =\frac{1}{4}$, in blue, $\gamma =\frac{5}{8}$.

\begin{figure}[H]
\begin{center}
\includegraphics[scale=0.68]{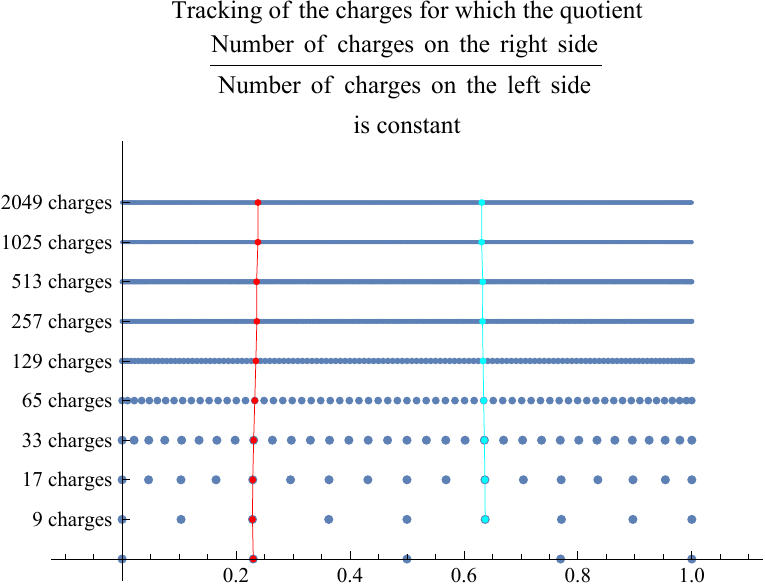}
\end{center}
\end{figure}

Here are their numerical values, showing that the sequences converge respectively to $\frac{1}{4}=0.25$ and $\frac{5}{8}=0.625$

\[ \begin{array}{|c|c|c|c|c|}
\hline
 \text{\textit{n}} & \frac{1}{4}(\text{\textit{$n$}}-1)+1 & \text{\textit{$x$}}_{\frac{1}{4}(\text{\textit{$n$}}-1)+1} & \frac{5}{8}(\text{\textit{$n$}}-1)+1
& x_{\frac{(n-1)}{4}+1} \\
\hline
 9 & 3 & 0.228829 & 6 & 0.637297 \\
\hline
 17 & 5 & 0.229322 & 11 & 0.636860 \\
\hline
 33 & 9 & 0.230887 & 21 & 0.635868 \\
\hline
 65 & 17 & 0.232830 & 41 & 0.634710 \\
\hline
 129 & 33 & 0.234755 & 81 & 0.633595 \\
\hline
 257 & 65 & 0.236486 & 161 & 0.632608 \\
\hline
 513 & 129 & 0.237976 & 321 & 0.631765 \\
\hline
 1025 & 257 & 0.239231 & 641 & 0.631058 \\
\hline
 \text{} & \text{} & \gamma = 0.25 &   & \gamma = 0.625 \\
\hline
\end{array} \]

As can be seen in the two examples in the figure and the table above, the point tracking curve having a fixed ratio $\frac{\text{Number of charges on the left}}{\text{Number of charges on the right}}$ converges to that fixed value when the number of loads increases. We will admit that this is always true.

Suppose the total charge on the needle is equal to $1$. If there are $n$ charges, this means that each charge has a value of $\frac{1}{n}$. If $\left( x_1, \ldots , x_n \right)$ is the list of positions of the charges on the needle (we assumed that it is $[0,1]$), this defines a discrete probability $\mathbb{P}_n$ on $[0,1]$, with $\mathbb{P}_n(A)=\frac{1}{n} \card \left[A\cap \left\{ x_i, 1 \leqslant i \leqslant n \right\} \right]$ for any Borelian $A$ of $[0,1]$. The distribution function $F_n$ of this probability is defined by 
\[ \text{for all } x \in \mathbb{R}, F_n(x)= \mathbb{P}_n (] - \infty , x]). \]

For all $n\in \llbracket 2, \infty \llbracket$, let us denote $\mathcal{R}_n$ the sequence giving the positions at equilibrium of charges when there are $2^n+1$ charges placed on the needle. Let us also denote $\mathcal{D}_2$ the set of dyadic numbers of $[0,1]$, that is to say numbers $\nu$ that can be written in the form $v=\frac{s}{2^i}$, with $i\in \llbracket 2, \infty \llbracket$, and $s\in \llbracket 0, 2^i \rrbracket$.\\
Let $u=\frac{q}{2^s}$ be a dyadic of $[0,1]$, and, for all $n\in \llbracket s+1, \infty \llbracket$, let $y_{u,n}$ be the position from the point of the equilibrium distribution to $2^n+1$ charges having a proportion of $\frac{q}{2^s}$ points to its left and $\frac{2^s-q}{2^s}$ points to its right. This is the $\left(\frac{q}{2^s}\times 2^n+1\right)$-th term of $\mathcal{R}_n$. By definition, $F_{2^n+1}\left(y_{u,n}\right)=u$. Furthermore, we saw that the sequence $\left(y_{u,n}\right)_{n \geqslant s+1}$ converges. Let us denote $z(u)$ its limit. This defines a function $z$ from $\mathcal{D}_2$ to $[0,1]$. We have $z(0)=0$, $z(1)=1$, and $z\left(\frac{1}{2}\right)=\frac{1}{2}$. The functions $F_n$ being increasing, so is $z$ which is also injective since $\mathcal{D}_2$ is dense in $[0,1]$. Again, since $\mathcal{D}_2$ is dense in $[0,1]$, and because the value of the jumps of $F_{2^n+1}$ tends towards $0$, $z$ admits a single continuous extension $\overline{z}$, from $[0,1]$ into $[0,1]$ which is a bijection. Let $G$ be the reciprocal function of $\overline{z}$. Then $G$ is the distribution function of a continuous law on $[0,1]$. Furthermore, for all $x\in [0,1]$, we have $\lim\limits_{n\to \infty } F_{2^n+1}(x)=G(x)$. We can also add that, according to Dini's theorem, there is uniform convergence of $\left(F_{2^n+1}\right)_{n\in \llbracket s+1, \infty \llbracket}$ towards $G$. In probability theory, one would say that there is {``}convergence in law{''}. Here is a graph showing the functions $F_{2^n+1}$, $n\in \llbracket 2, 11 \rrbracket$ (the vertical segments of the graphs have been added to get a better readability).

\begin{prop}
The sequence $\left(F_{2^n+1}\right)_{n\in \mathbb{N}^*}$ converges to $G : x \mapsto x$, the distribution function of the uniform density.
\end{prop}

\begin{figure}[H]
\begin{center}
\includegraphics[scale=0.7]{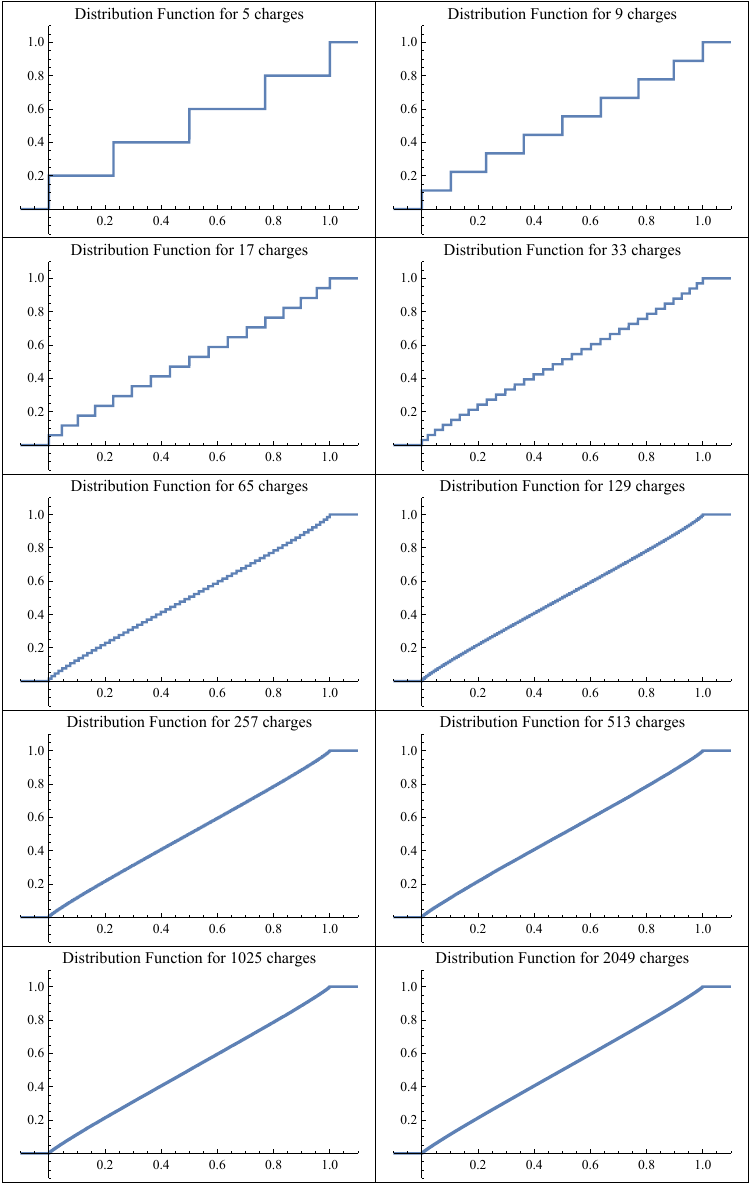}
\end{center}
\end{figure}

Before seeing the mathematical consequences, let us note a few points:

\textbullet\ In the graphs of the first distribution functions, we observe that the {``}steps{''} in the center of the graph are a little longer than those located at the ends. Once again, this confirms a remark by Griffiths and Li in~\cite{griffiths}: {``}It seems obvious that Coulomb repulsion will push charge out towards the ends{''}... {``}presumably some of it is left spread out along the length of the needle{''}. To support this remark, here is a graph giving the ratios of lengths between the largest of the intervals separating two consecutive charges (at the center of the needle), and the smallest (at the end), as a function of the number of charges.

\begin{figure}[H]
\begin{center}
\includegraphics[scale=0.65]{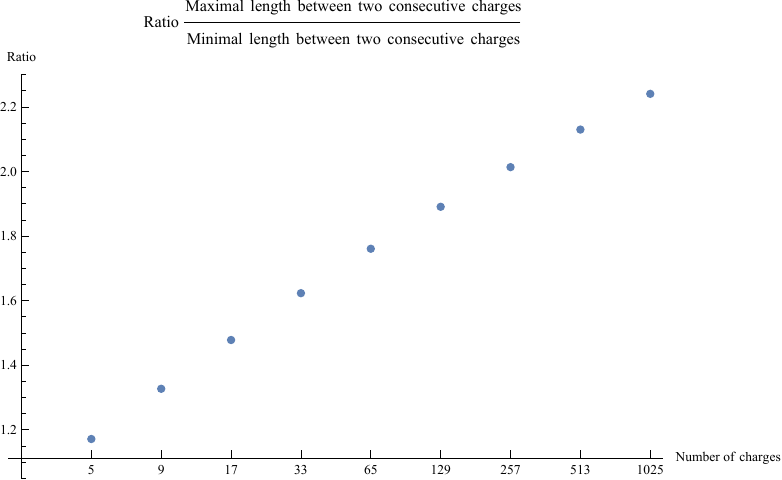}
\end{center}
\end{figure}

\textbullet\ When the number of loads increases, this difference between the {``}step heights{''} diminishes. We can see it in the previous graph: the red line and the blue line are not completely straight. This is due to the rebalancing of loads towards a uniform arrangement.

\textbullet\ The convergence of the distribution functions towards a continuous uniform distribution confirms the results of Djordjevi{\' c} (1985), Good (1997), Andrews (1997) and those of Amir and Matzner (2004).

\section*{6. Back to the differential systems}

We saw in section 1 that when placed on a needle, $n$ charges move indefinitely around the equilibrium position. In the simulations of section 1, the initial condition was the equirepartition of the charges on the needle. In that situation, the move around the equilibrium was reduced. What if the charges are placed in another initial condition? In that case, oscillations around the equilibrium position might be more important. Would that change the convergence towards the uniform distribution?

In order to answer that question, we have studied the differential system with an initial condition where the charges are concentrated on one half of the needle: $x_i(0)=\frac{i-1}{2n-1}$ if $1<i<n$, $x_n(0)=1$. Below, the first graphics shows the trajectories of $17$ charges on the interval $[0,3]$. Indeed, they have large amplitudes. However, as shown in the three next graphics, corresponding to time $1$, $2$ and $3$, these large amplitudes do not imply large variations around the identity of the distribution function. And as show the next graphics for $65$ charges, whatever the initial conditions, the distribution function converges towards the distribution function of the uniform density when the number of charges increases. That means that the analysis of section 5 remains true when taking into account the fact that the charges do not stay at a fixed place.

\begin{figure}[H]
\begin{center}
\includegraphics[scale=0.65]{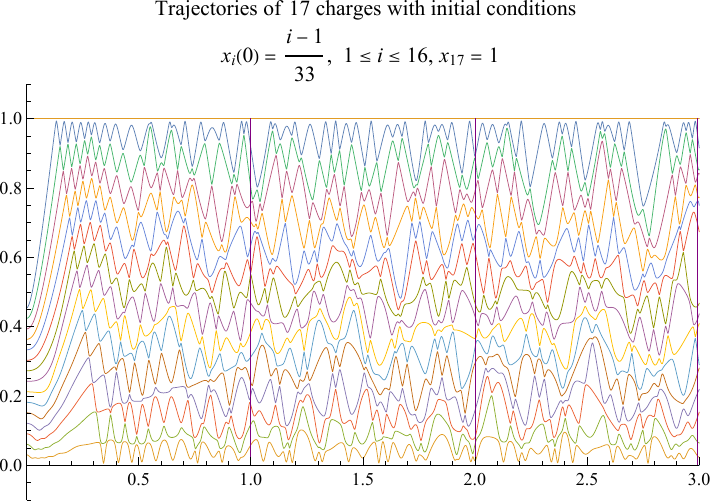}
\end{center}
\end{figure}

\begin{figure}[H]
\begin{center}
\includegraphics[scale=0.65]{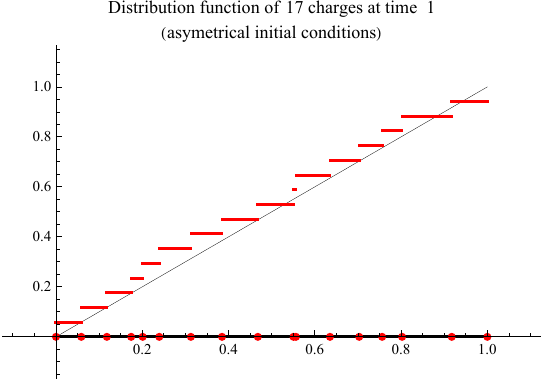}
\end{center}
\end{figure}

\begin{figure}[H]
\begin{center}
\includegraphics[scale=0.65]{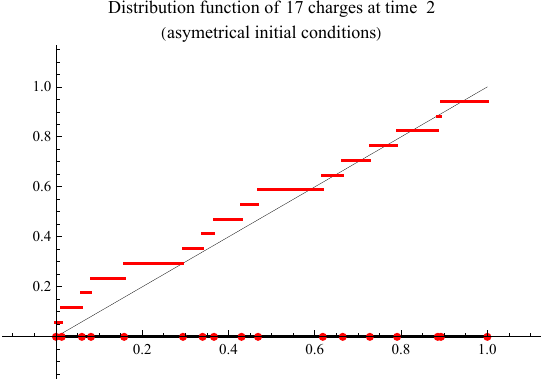}
\end{center}
\end{figure}

\begin{figure}[H]
\begin{center}
\includegraphics[scale=0.65]{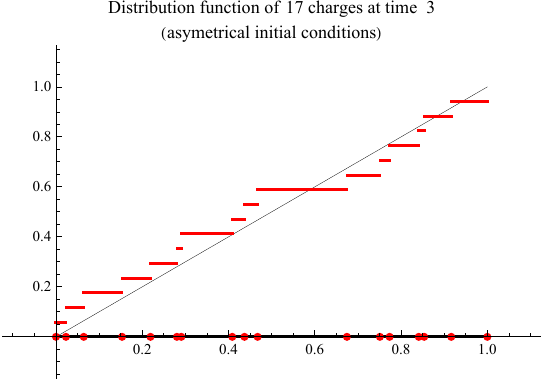}
\end{center}
\end{figure}

\begin{figure}[H]
\begin{center}
\includegraphics[scale=0.5]{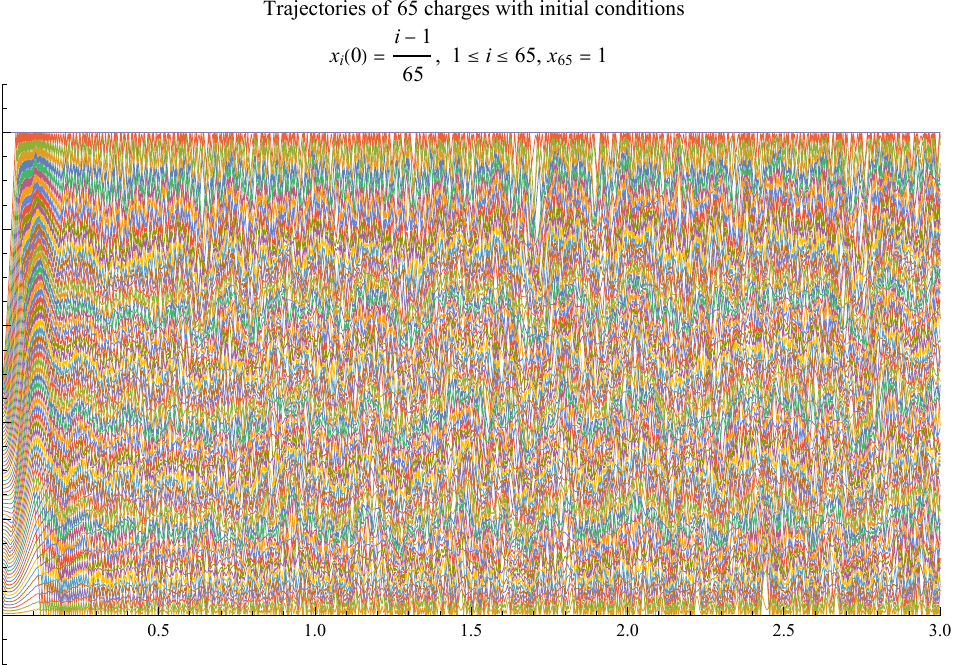}
\end{center}
\end{figure}

\begin{figure}[H]
\begin{center}
\includegraphics[scale=0.65]{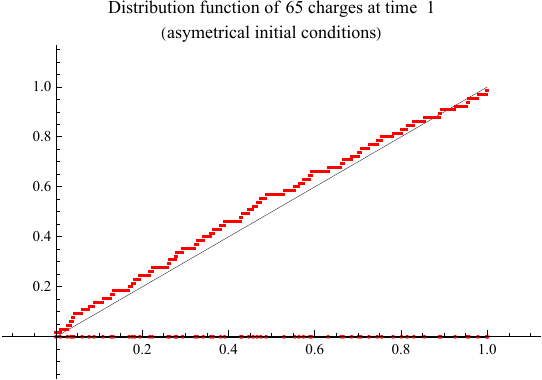}
\end{center}
\end{figure}

\begin{figure}[H]
\begin{center}
\includegraphics[scale=0.65]{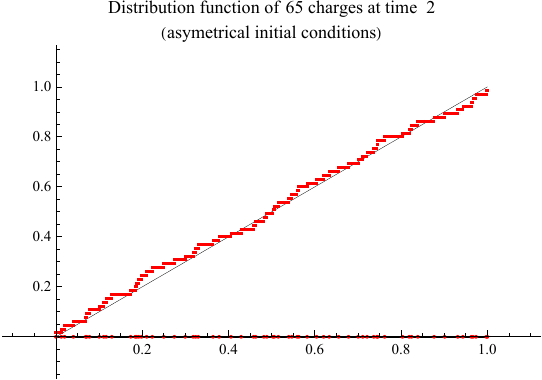}
\end{center}
\end{figure}

\begin{figure}[H]
\begin{center}
\includegraphics[scale=0.65]{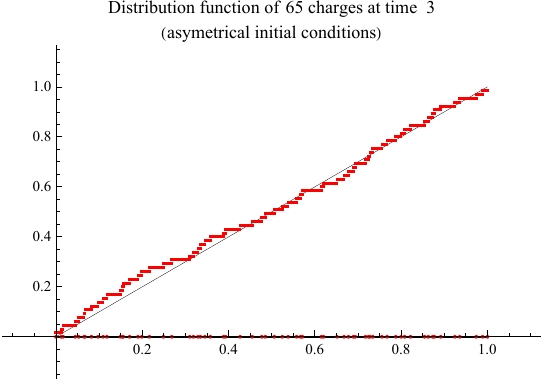}
\end{center}
\end{figure}

\section*{7. Consequences}

\subsection*{7.1 --- Field produced by the charges on the needle}

Let us now return to the mathematical consequences of the evolution of the distribution of charges that we observed.

These experimental results show that on $[0,1]$, $\left( F_{2^n+1} \right)_{n\in \mathbb{N}^*}$ simply converges to $G: \left\lbrace
  \begin{array}{rcl}
    [0,1] & \to & [0,1] \\
    x & \mapsto & x     \\
  \end{array}
\right.$. Due to the properties of convexity and monotony established in section 4, we can infer that $\left( F_n \right)_{n\in \mathbb{N}^*}$ simply converges to $G: \left\lbrace
  \begin{array}{rcl}
    [0,1] & \to & [0,1] \\
    x & \mapsto & x     \\
  \end{array}
\right.$. We can therefore use the theory of Riemann-Stieltjes integral. Consider for example the field created at a point $X\in \mathbb{R}^3$ by $n$ charges placed on the needle in the equilibrium position. Let us always denote $\mathcal{R}_n$ the sequence giving the positions on $[0,1]$ at the equilibrium of these charges, $\mathcal{R}_n=\left( x_1, \ldots , x_n \right)$. In $\mathbb{R}^3$, the charge having abscissa $x_i$ therefore has coordinates $X_i= \left( x_i, 0, 0 \right)$. This field $\mathcal{C}(X)$ is given (up to multiplicative constants) by 
\[ C(X) = \frac{1}{n} \sum_{i=1}^n \frac{1}{\left\| X-X_i \right\|^2} \times \frac{X - X_i}{\left\| X-X_i \right\|} . \]

Thanks to the Riemann-Stieltjes integral, this field can be expressed in the form 
\[ C(X) = \sideset{_{\mathcal{R-S}}\,}{_0^1}\int \frac{1}{\left\| X - (t,0,0) \right\|^2} \times \frac{X - (t,0,0)}{\left\| X - (t,0,0) \right\|} \mathrm{d}F_n (t). \]

We can therefore apply the following convergence theorem (Gordon p. 200,~\cite{gordon}):

\begin{theo}
Let $f$ a continuous function defined on $[a,b]$ ($a<b$), and $( \varphi_n)_{n \in \mathbb{N}}$ be a sequence of functions of bounded variation on $[a,b]$ and suppose that $( \varphi_n)_{n \in \mathbb{N}}$ converge to $\varphi$ pointwise on $[a,b]$. Suppose that there exists a (strictly) positive number $M$ such that for all $n$, $V\left(\varphi_n, [a,b]\right) \leqslant M$ (where $V\left(\varphi_n, [a,b]\right)$ is the variation of $\varphi_n$ on $[a,b]$). Then, 
\[ \sideset{_{\mathcal{R-S}}\,}{_a^b}\int f \mathrm{d} \varphi = \lim\limits_{n\to \infty} \sideset{_{\mathcal{R-S}}\,}{_a^b}\int f \mathrm{d} \varphi_n . \]
\end{theo}

Here, we are dealing with distribution functions and their variation is equal to $1$. We also have pointwise convergence of $\left( F_n \right)_{n\in \mathbb{N}^*}$ towards $G$. If $X\notin [0,1]\times \{0\}\times \{0\}$ (the needle), the function $\left\lbrace
  \begin{array}{rcl}
    [0,1] & \to & \mathbb{R}^3 \\
    t & \mapsto & \frac{1}{\left\| X - (t,0,0) \right\|^2} \times \frac{X - (t,0,0)}{\left\| X - (t,0,0) \right\|}     \\
  \end{array}
\right.$ is defined on $[0,1]$. The theorem therefore applies and 

\noindent \resizebox{\linewidth}{!}{$\sideset{_{\mathcal{R-S}}\,}{_0^1}\int \frac{1}{\left\| X - (t,0,0) \right\|^2} \times \frac{X - (t,0,0)}{\left\| X - (t,0,0) \right\|} \mathrm{d} t = \lim\limits_{n\to \infty} \sideset{_{\mathcal{R-S}}\,}{_0^1}\int \frac{1}{\left\| X - (t,0,0) \right\|^2} \times \frac{X - (t,0,0)}{\left\| X - (t,0,0) \right\|} \mathrm{d} F_n (t) .$}

Note that due to the result of section 6, the convergence still apply (although more slowly) when we considere that the charges move on the needle.

Theorem 1 explains how the convergence towards the continuous uniform distribution has to be understood :

\begin{csq}
The convergence is the pointwise convergence of the sequence of the distribution functions associated to the charges on the needle towards the distribution function of the continuous uniform distribution.
\end{csq}

\begin{csq}
The convergence of the distribution functions is not affected by the fact that some charges concentrate towards the ends of the needle.
\end{csq}

\begin{csq}
The convergence of the distribution functions is not affected by the fact that the charges move constantly.
\end{csq}

\begin{csq}
Theorem 1 shows that when the number of charges increases, it is legitimate to approximate the field created by the charges on the needle by the field generated by a continuous uniform distribution.
\end{csq}

\begin{csq}
Theorem 1 provides a kind of {``}equivalence principle{''} : in order to study the field created by a family of distributions of charges, one can use any family of distributions of charges, as far as the distributons functions of both families converge to the same distribution function.
\end{csq}

These consequences are visually illustrated in the accompanying animated short film available at the link \url{https://sdrive.cnrs.fr/s/TT45z7LFbXdPwHn?dir=undefined&openfile=332835629}.

If $G$ is the identity function, the Riemann-Stieltjes integral coincides with the Riemann integral. As the latter is the usual integral, we will write it $\int \cdot \, dx$ instead of $\sideset{_{\mathcal{R-S}}\,}{}\int \cdot \, dx$. So we have 
\resizebox{\linewidth}{!}{$\begin{array}{lcl}
\sideset{_{\mathcal{R-S}}\,}{_0^1}\int \frac{1}{\left\| X - (t,0,0) \right\|^2} \times \frac{X - (t,0,0)}{\left\| X - (t,0,0) \right\|} \mathrm{d} t &= &\int_0^1 \frac{1}{\left\| X - (t,0,0) \right\|^2} \times \frac{X - (t,0,0)}{\left\| X - (t,0,0) \right\|} \mathrm{d} t \\
&= &\lim\limits_{n\to \infty} \sideset{_{\mathcal{R-S}}\,}{_0^1}\int \frac{1}{\left\| X - (t,0,0) \right\|^2} \times \frac{X - (t,0,0)}{\left\| X - (t,0,0) \right\|} \mathrm{d} F_n (t) .
\end{array}$}

When $X=(x,0,0)\in [0,1]\times \{0\}\times \{0\}$, that is to say if $X$ is a point of the needle, the term $\frac{1}{\| X-(t,0,0)\|^2}\times \frac{X-(t,0,0)}{\| X-(t,0,0)\|}$ becomes $\frac{\sgn (x-t)}{(x-t)^2}$. This function of $t$ is not defined at $x$, so the theorem does not apply.

However, the integral can be calculated using Cauchy's idea of principal part. Indeed, if $x\in [0,1]$, if $\varepsilon >0$, if $x-\varepsilon >0$ and if $x+\varepsilon <1$, \\

\noindent\resizebox{\linewidth}{!}{$\int_0^{x-\varepsilon } \frac{1}{(x-t)^2} \mathrm{d} t - \int_{x+\varepsilon}^1 \frac{1}{(x-t)^2} \mathrm{d} t = \frac{1}{x - 1}-\frac{1}{x}=\frac{2 x - 1}{x(x-1)}.$}\\

This does not depend on $\varepsilon$, and we can therefore write, for a continuous total charge density equal to 1 (still not taking into account the multiplicative constants), that the field at $x\in  [0,1]$ is given by

\[ \mathcal{C}(x)=\frac{2 x - 1}{x(x-1)} . \]

Here is the representation of this field

\begin{figure}[H]
\begin{center}
\includegraphics[scale=0.65]{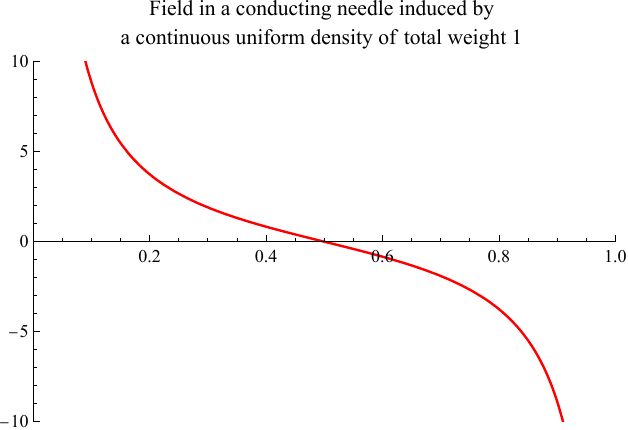}
\end{center}
\end{figure}

Note that if we consider a constant sign, we have 
\[ \int_0^{x-\varepsilon} \frac{1}{(x-t)^2} \mathrm{d} t+\int_{x+\varepsilon}^1 \frac{1}{(x-t)^2} \mathrm{d} t=\frac{1}{-1+x}-\frac{1}{x}+\frac{2}{\varepsilon} \]

which tends towards $\infty$ when $\varepsilon$ tends towards $0$. The integral diverges (whatever the theory of the integral considered). This explains the note at the beginning of this paper, in the subsection {``}motivations{''} of the section {``}Notations{''}.

\subsection*{7.2 --- Application of the equivalence principle --- On the needle, comparison of the discrete field and the continuous field with equal distribution (see Section VI of \cite{andrews})}

We explained above that the charges on a needle do not have a uniform distribution. However, when the number of charges increase, both the repartitions of the charges and the uniform discrete repartitions converge in the way explained above (of the distribution functions) towards the uniform continuous distribution. We can apply the equivalence principle, and approximate the field created by the charges on the needle by the field created by a virtual family of charges which would place themselves according to the uniform discrete distribution. Here are some formulas obtained

\textbullet\ The closed formulas of this section were provided by Mathematica. What is the relative magnitude of the forces exerted on a charge by the nearest charges? Given the convergence to a uniform distribution, let's do the calculations assuming the points are equally distributed along the needle; this allows an exact formal calculation and provides a conforming asymptotic result, due to convergence towards a uniform distribution. Consider again $u=\frac{q}{2^s}$ a dyadic of $[0,1]$, $n\in \llbracket s+1, \infty \llbracket$; we assume that the $2^n+1$ charges are equally distributed on the needle. Let $C_{u, n}$ be the charge having a proportion of $\frac{q}{2^s}$ points to its left and $\frac{2^s-q}{2^s}$ points to its right. This charge is placed at the point of abscissa $\frac{q}{2^s}$. The ratio $Q_{q,n}^-$ of the force exerted on $C_{q, n}$ by the charge immediately to its left to the total force exerted on $C_{q, n}$ by the charges of $\left[0, \frac{q}{2^s}\right[$ is given by 
\resizebox{\linewidth}{!}{$Q_{q,n}^- = \frac{\frac{1}{(2^n)^2}}{\sum\limits_{k=0}^{q 2^{n-s} - 1} \frac{1}{\left( \frac{k}{2^n} - \frac{q}{2^s} \right)^2}} = \frac{3 \times 2^{1-4n}}{\pi^2 - 6 \PolyGamma \left[ 1, 1+2^{n-s} q \right]}$}

and we have $\lim\limits_{n\to \infty } Q_{q, n}^-=0$.

Likewise, the ratio $Q_{q,n}^+$ of the force exerted on $C_{q, n}$ by the charge immediately to its right to the total force exerted on $C_{q,n}$ by the charges of $\left] \frac{q}{2^s}, 1\right]$ is given by

\noindent \resizebox{\linewidth}{!}{$Q_{q,n}^+ = \frac{\frac{1}{(2^n)^2}}{\sum\limits_{k=q 2^{n-s} +1}^{2^n} \frac{1}{\left( \frac{k}{2^n} - \frac{q}{2^s} \right)^2}} = \frac{3 \times 2^{1-4n}}{\pi^2 - 6 \PolyGamma \left[ 1, 1+ 2^n +2^{n-s} q \right]}$}

and we have $\lim\limits_{n\to \infty } Q_{q, n}^+=0$.

\textbullet\ When we normalize the total charge to $1$, the charge on $\left[ 0, \frac{q}{2^s} \right[$ is equal to $\frac{q}{2^s}$, and does not depend on the number of charges present on this interval. One might think that the force exerted on $C_{u, n}$ by these charges converges, since the total charge is constant. It is not so. Indeed, 

\noindent \resizebox{\linewidth}{!}{$\frac{1}{2^n + 1} \sum\limits_{k=0}^{q 2^{n-s} - 1} \frac{1}{\left( \frac{k}{2^n} - \frac{q}{2^s} \right)^2} = \frac{2^{-1+2n} \left( \pi^2 - 6 \PolyGamma \left[ 1,1+2^{n-s} q \right] \right)}{3(1+2^n)}$}

and this term tends to $\infty$ when $n$ tends to $\infty$. The same is true of the force exerted on $C_{u, n}$ by the charges placed on $\left] \frac{q}{2^s} , 1 \right]$.

\textbullet\ Now consider the total field created on $C_{u, n}$ by the other $2^n$ charges. We are still interested in the case of equal distribution. We obtain 

$\begin{array}{cl}
&\frac{1}{(2^n + 1)} \left( \sum\limits_{k=q 2^{n-s} +1}^{2^n} \frac{1}{\left( \frac{k}{2^n} - \frac{q}{2^s} \right)^2} - \sum\limits_{k=0}^{q 2^{n-s} - 1} \frac{1}{\left( \frac{k}{2^n} - \frac{q}{2^s} \right)^2} \right) \\
= &\frac{4^n \left(- \PolyGamma \left[ 1,1+2^n - 2^{n-s}q \right] + \PolyGamma \left[ 1,1+ 2^{n-s}q \right] \right)}{1+2^n} .
\end{array}$

By noting $x=\frac{q}{2^s}\in [0,1]$, we have 

\noindent \resizebox{\linewidth}{!}{$\lim\limits_{n\to \infty} \frac{4^n \left(- \PolyGamma \left[ 1,1+2^n - 2^{n-s}q \right] + \PolyGamma \left[ 1,1+ 2^{n-s}q \right] \right)}{1+2^n} = \frac{1}{x} - \frac{1}{1-x} .$}

We saw above that we obtain the same thing with an equally distributed diffuse charge.

\subsection*{7.3 --- Out of the needle, comparison of the field induced by a finite distribution of charges with the field induced by a continuous density of charges}

In that subsection, we considere the field induced by a finite set of charges which are assumed to be in a fixed position, at equilibrium. 

For this comparison, we have considered two parameters :

\textbullet\ Comparison of the fields obtained for a fixed number of charges, as a function of the distance to the needle,

\textbullet\ Comparison of the fields obtained at a fixed distance of the needle, as a function of the number of charges on the needle. 

As we saw above, consider for example the field created at a point $X\in \mathbb{R}^3$ by $2^n+1$ charges placed on the needle in the equilibrium position. Let us always denote $\mathcal{R}_n$ the sequence giving the positions on $[0,1]$ at the equilibrium of these charges, $\mathcal{R}_n=\left( x_1, \ldots , x_{2^n+1} \right)$. In $\mathbb{R}^3$, the charge having abscissa $x_i$ therefore has coordinates $X_i= \left( x_i, 0, 0 \right)$.
This field $\mathcal{C}_n(X)$ is given (up to multiplicative constants) by 
\[ \mathcal{C}_n (X) = \frac{1}{2^n + 1}  \sum_{i=1}^{2^n + 1} \frac{1}{\left\| X - X_i \right\|^2} \times \frac{X - X_i}{\left\| X - X_i \right\|} . \]

For a uniform density, the field in $X$ is given by 
\[ \mathcal{C} (X) = \int_0^1 \frac{1}{\left\| X - (t,0,0) \right\|^2} \times \frac{X - (t,0,0)}{\left\| X - (t,0,0) \right\|} \mathrm{d} t. \]

In order to compare these fields, we represented them for points at a distance $\rho$ from the needle, and we varied $n$ and $\rho$. The problem having cylindrical symmetry, the points $X$ belong to the coordinate plane of equation $z=0$ in a coordinate system $(O,x,y,z)$. Below are some of the resulting graphs. For visibility, the field was chosen {``}outgoing{''}.

\begin{figure}[H]
\begin{center}
\includegraphics[scale=0.6]{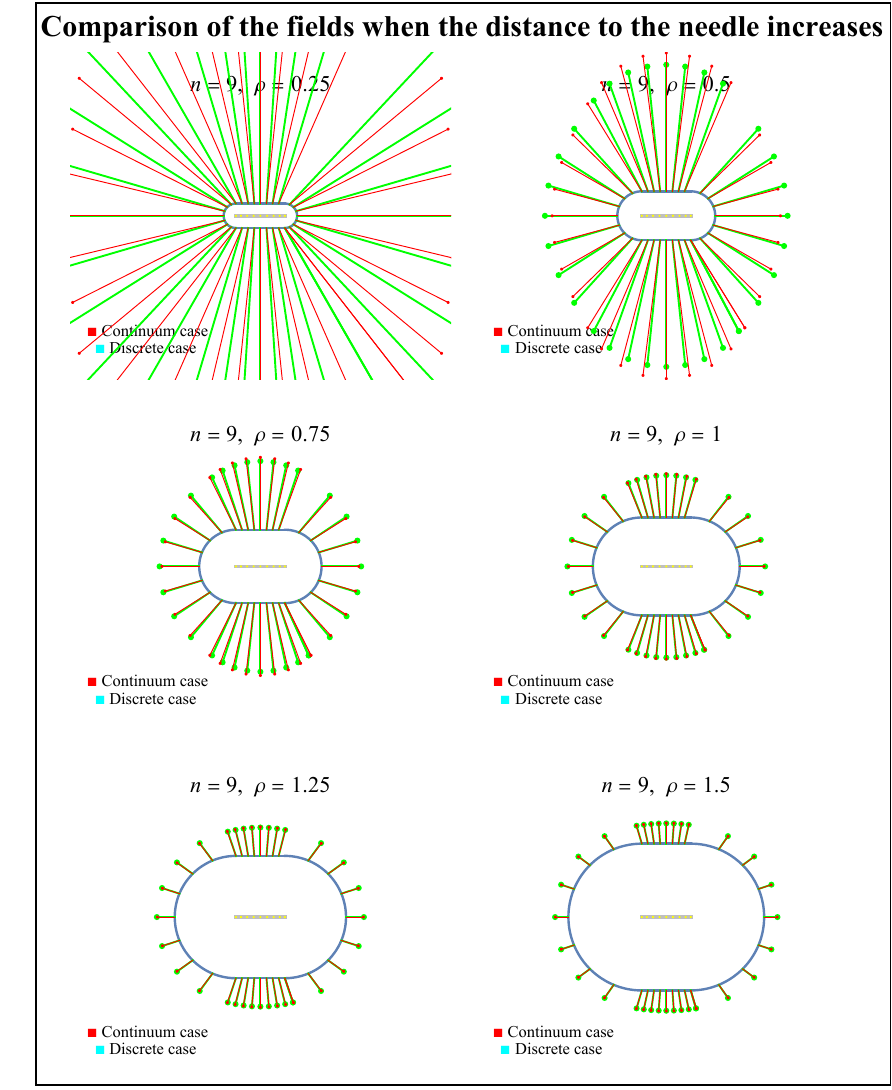}
\end{center}
\end{figure}

\begin{figure}[H]
\begin{center}
\includegraphics[scale=0.6]{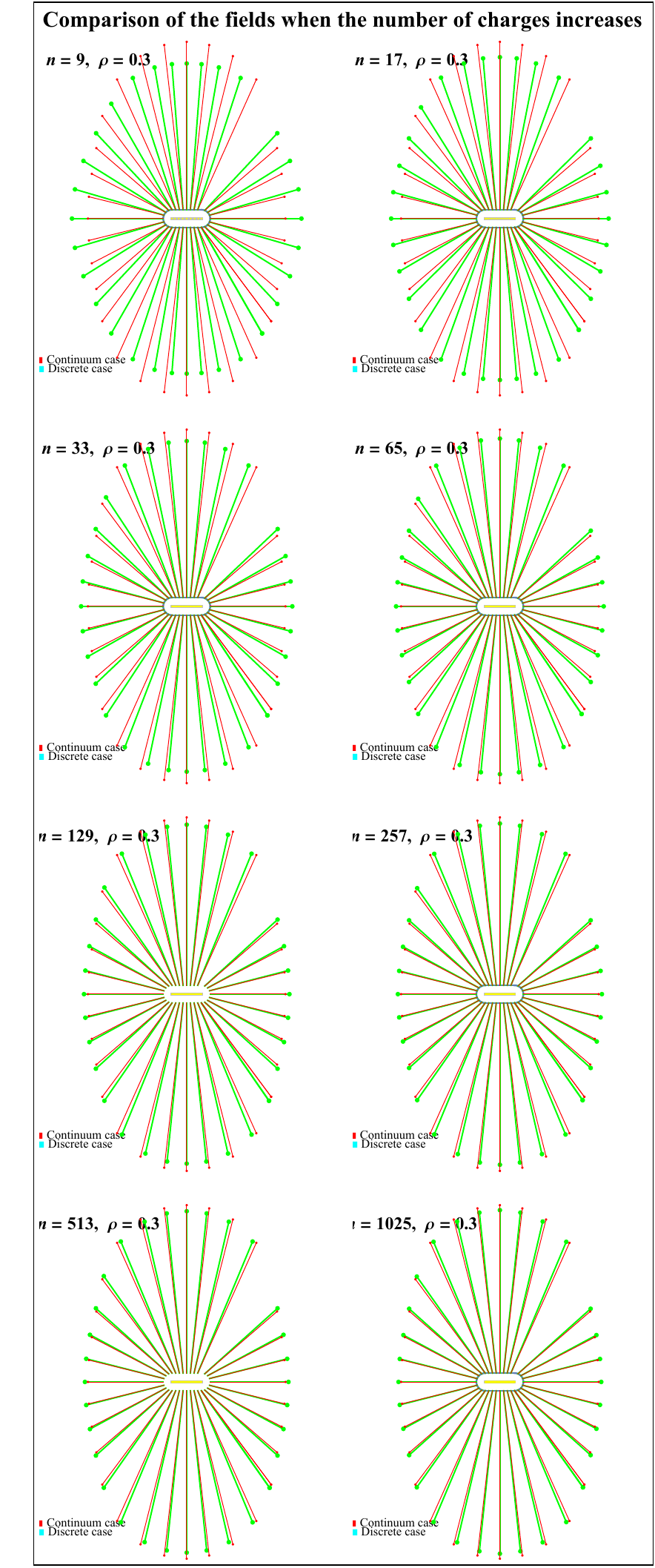}
\end{center}
\end{figure}

Three remarks on these figures:

\textbullet\ As expected, the length of the arrows, i.e. the field intensity, decreases with distance. Since the total charge is always the same, this is not the case when the number of charges increases.

\textbullet\ The number of charges being fixed, the closer we get to the needle, the more the fields due to these charges differ from the field created by a continuous density of charges.

\textbullet\ The distance remaining the same, the more the number of charges increases, the more the two fields tend to align.

\section*{Acknowledgements}

The research of O. Ciftja was supported in part by
National Science Foundation (NSF) grant no. DMR-2001980.

\end{document}